\documentclass{jfm}

\usepackage{amssymb,amsfonts,amsmath}
\usepackage{graphicx}
\usepackage{natbib}
\usepackage{color}

\graphicspath{
  {/Users/fberon/Documents/Latex/Pub/Personales/LagDyn/BlkHoles/Figures/}
  {/Users/fberon/Documents/Latex/Pub/Personales/LagDyn/BlkHoles/Figures/blackholes-fig03/}
}

\def\cf{\emph{cf.} }

\title{Addendum to ``Coherent Lagrangian vortices: The black holes
of turbulence''}

\author[G. Haller and F. J. Beron-Vera]{G.\ Haller$^1$\thanks{Email
address for correspondence: georgehaller@ethz.ch}\ns and F.\ J.\
Beron-Vera$^2$}

\affiliation{$^1$Institute for Mechanical Systems, ETH Zurich,
Zurich, Switzerland\\[\affilskip] $^2$ Department of Atmospheric
Sciences, RSMAS, University of Miami, Miami FL, USA}

\pubyear{2013}
\volume{X}
\pagerange{X--X}
\date{20 May 2014; revised 20 May 2014; accepted 20
May 2014.}
\begin{document}

\maketitle

\begin{abstract}
  In \citet{Haller-Beron-13} we developed a variational principle
  for the detection of coherent Lagrangian vortex boundaries. The
  solutions of this variational principle turn out to be closed
  null-geodesics of the Lorentzian metric induced by a generalized
  Green--Lagrange strain tensor family. This metric interpretation
  implies a mathematical analogy between coherent Lagrangian vortex
  boundaries and photon spheres in general relativity. Here we give
  an improved discussion on this analogy.
\end{abstract}

\section{The main results of \citet{Haller-Beron-13}}

We consider a two-dimensional velocity field $v(x,t)$, with $x$
labeling the location within a two-dimensional region $U$ of interest
and with $t$ referring to time. Fluid trajectories generated by
$v(x,t)$ are denoted $x(t;t_{0},x_{0}$), with $x_{0}$ referring to
the initial position of the trajectory at time $t_{0}$. These
trajectories solve the differential equation
\begin{equation}
  \dot{x} = v(x,t),
  \label{eq:uxt}
\end{equation}
and generate the flow map 
\begin{equation}
  F_{t_{0}}^{t}(x_{0}) := x(t;t_{0},x_{0}),
\end{equation}
which takes an initial position $x_{0}$ at time $t_{0}$ to its
current position at time $t$.

The right Cauchy--Green strain tensor field associated with the
flow map is defined as $C_{t_{0}}^{t}(x_{0}) = \nabla
F_{t_{0}}^{t}(x_{0})^\top \nabla F_{t_{0}}^{t}(x_{0})$, with
eigenvalues $\lambda_{i}(x_{0})$ and eigenvectors $\xi_{i}(x_{0})$
satisfying
\begin{equation}
C_{t_{0}}^{t}\xi_{i} = \lambda_{i}\xi_{i},\quad
\left|\xi_{i}\right| = 1,\quad i=1,2;\quad
0<\lambda_{1}\leq\lambda_{2},\quad
\xi_{1}\perp\xi_{2}.
\end{equation}
In \citet{Haller-Beron-13} we sought the time $t_{0}$ positions
of Lagrangian vortex boundaries as closed stationary curves of the
averaged Lagrangian strain. Such curves turned out to coincide with
the zero-energy solutions of a one-parameter family of variational
problems defined as
\begin{equation}
  \delta\mathcal{E}_{\lambda}(\gamma)=0,\quad 
  \mathcal{E}_{\lambda}(\gamma)
  = \oint_{\gamma}\langle r^{\prime}(s),
  E_{\lambda}(r(s))r^{\prime}(s)\rangle\,\mathrm{d}s,\quad
  \lambda\in\mathbb{R}^{+}.
  \label{eq:qdef-1}
\end{equation}
Here the strain energy functional $\mathcal{E}_{\lambda}(\gamma)$
is defined through the generalized Green--Lagrange strain tensor
family
\begin{equation}
  E_{\lambda}(x_{0}) = \frac{1}{2}[C_{t_{0}}^{t}(x_{0})-\lambda^{2}I].
  \label{eq:Elambda}
\end{equation}

Consider the flow domain
\begin{equation}
  U_{\lambda} = \{x_0 \in U \mid \lambda_1(x_0) < \lambda^2
  < \lambda_2(x_0)\},
  \label{eq:Ulda}
\end{equation}
where the tensor field $E_{\lambda}$ has two nonzero eigenvalues
of opposite sign.  Then the quadratic function
\begin{equation}
  g_{\lambda}(u,u) = \left\langle u,E_{\lambda}u\right\rangle 
  \label{eq:g}
\end{equation}
defines a Lorentzian metric \citep{Beem-etal-96} on $U_{\lambda}$,
with signature $(-,+)$ inherited from the eigenvalue configuration
of $E_{\lambda}$. The zero-energy solutions of (\ref{eq:qdef-1})
are therefore precisely the closed null-geodesics of the Lorentzian
metric $g_{\lambda}$, which satisfy one of the two differential
equations
\begin{equation}
   r^{\prime}(s) = \eta_{\lambda}^{\pm}(r(s)),\quad
	\eta_{\lambda}^{\pm}(r) = 
	\sqrt{\frac{\lambda_{2}(r)-\lambda^{2}}{\lambda_{2}(r)-\lambda_{1}(r)}}\,\xi_{1}(r)\pm
	\sqrt{\frac{\lambda^{2}-\lambda_{1}(r)}{\lambda_{2}(r)-\lambda_{1}(r)}}\,\xi_{2}(r).
	\label{eq:ode-1}
\end{equation}
In \citet{Haller-Beron-13} we concluded that closed orbits of
(\ref{eq:ode-1}) (termed \emph{closed $\lambda$-lines}) must
necessarily encircle metric singularities of $g_{\lambda}$. Such
singularities occur at points $x_{0}$ where $\lambda_{1}(r)=\lambda_{2}(r)$
holds for the eigenvalues of the Cauchy--Green strain tensor.

The Lorentzian metric interpretation discussed above implies a
geometric analogy between coherent Lagrangian vortex boundaries and
photon spheres in cosmology. Below we give more detail on this
analogy, followed by an improved version of its summary with a more
relevant reference.

\section{More on the analogy with photon spheres}

In the vicinity of any $\lambda$-line in $U_{\lambda}$, the vector
fields $\xi_{i}$ define a curvilinear coordinate system with pointwise
orthogonal coordinate lines. Direct substitution of $\xi_{i}$ into
the metric (\ref{eq:g}) gives
\begin{equation}
  g_{\lambda}(\xi_{i},\xi_{i}) = \frac{1}{2}(\lambda_{i} -
  \lambda^{2}),
\label{eq:xi_sign}
\end{equation}
showing that $g_{\lambda}(\xi_{1},\xi_{1})<0$ and
$g_{\lambda}(\xi_{2},\xi_{2})>0$ everywhere in $U_\lambda$.  This
shows that $\xi_{1}$-trajectories form the time-like coordinates
and the $\xi_{2}$-trajectories form the space-like coordinates of
the metric $g_{\lambda}$ in $U_{\lambda}$ \citep{Beem-etal-96}. We
further note that
$g_{\lambda}(\eta_{\lambda}^{\pm},\eta_{\lambda}^{\pm})=0$, and
hence a $\lambda$-line is \emph{nowhere space-like} in the language
of Lorentzian geometry. Given that our closed null-geodesics are
nowhere space-like hypersurfaces built out of null-geodesics, they
are \emph{photon surfaces} by the general definition of
\citet{Claudel-etal-01}.

\begin{figure}
  \centering%
  \includegraphics[width=.7\textwidth]{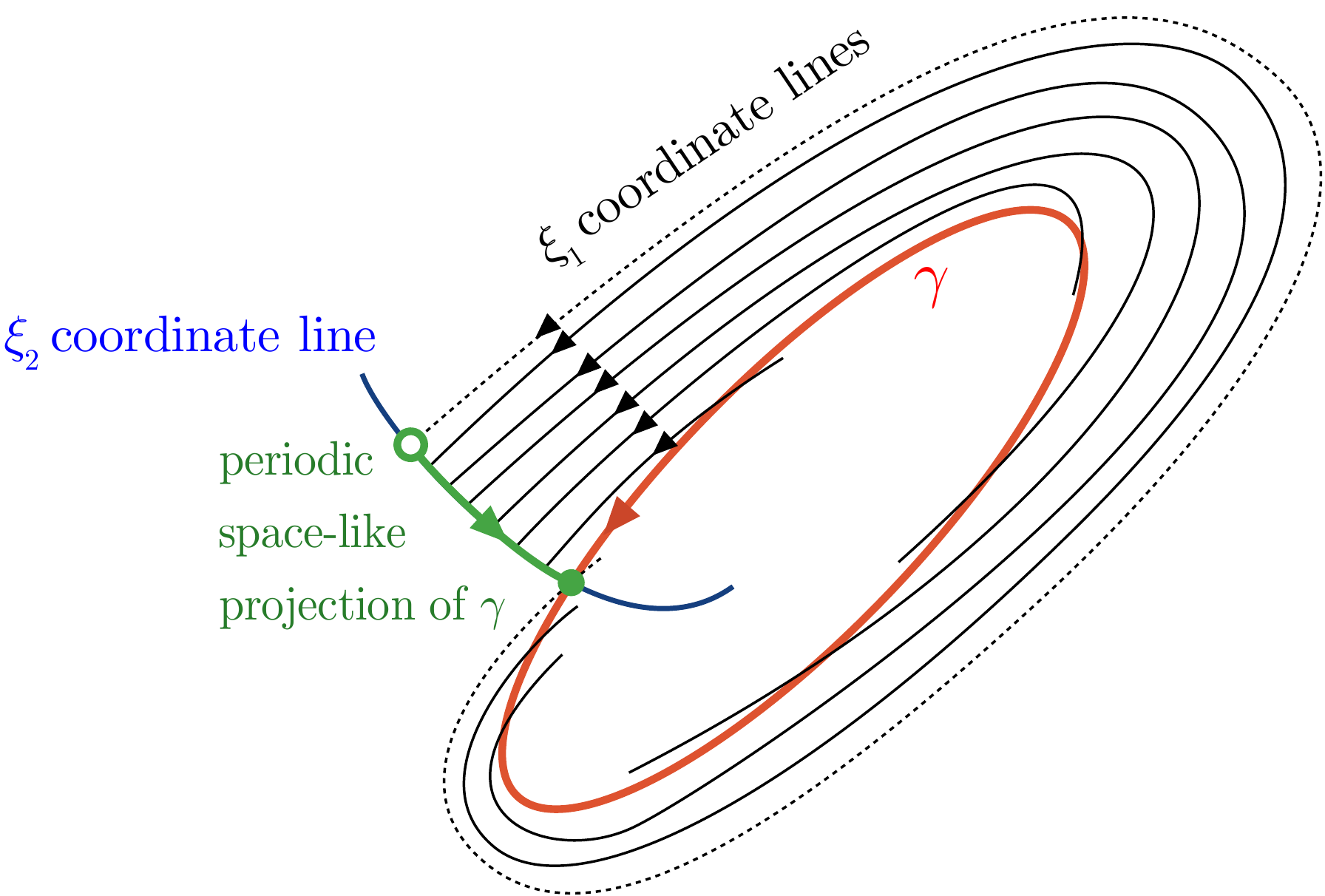}%
  \caption{A closed null-geodesic $\gamma$ of the Lorentizan metric
  $g_{\gamma}$ has a periodic space-like projection along the
  time-like coordinate lines. This space-like projection has a
  jump-discontinuity due to the low dimensionality of the space-time,
  but evolves periodically along with the underlying closed
  null-geodesic. This is in contrast to general $\lambda$-lines
  that have aperiodic space-like projections.\label{fig:proj}}
\end{figure}

Next we note that 
\begin{equation}
  \left\langle \eta_{\lambda}^{\pm}(r),\xi_{1}(r)\right\rangle =
  \sqrt{\frac{\lambda_{2}(r)-\lambda^{2}}{\lambda_{2}(r)-\lambda_{1}(r)}}\in(0,1),\quad 
  r\in U_{\lambda}.
\end{equation}
Consequently, trajectories of the $\xi_{1}(r)$ line field ($\xi_{1}$
coordinate lines) intersect any closed $\lambda$-line $\gamma$
transversally, with an angle of intersection that is always less
than $\frac{1}{2}\pi$, as sketched qualitatively in Figure \ref{fig:proj}.
In the same figure, we also show a representative trajectory of the
$\xi_{2}(r)$ line field (a $\xi_{2}$ coordinate line) which is
pointwise orthogonal to the $\xi_{1}$ coordinate lines by construction.
We conclude that the projection of $\gamma$ onto a space-like
submanifold along the time-like coordinates results in a periodic
(albeit discontinuous) space-like orbit (\cf Figure
\ref{fig:proj}).

In summary, a closed orbit $\gamma$ of the $\eta_{\lambda}^{\pm}(r)$
vector field is a photon surface of the $(U_{\lambda},g_{\lambda})$
space-time. Geodesics forming this photon surface have
periodically moving projections on the space-like coordinates, with
the projection taken along the time-like coordinate lines.

The orbit $\gamma$, therefore, satisfies a plausible extension of
the Claudel--Virbhadra--Ellis definition of a \emph{photon sphere}
\citep{Claudel-etal-01} from symmetric higher-dimensional space-times
to non-symmetric two-dimensional space-times. This extension relaxes
the requirement for a rotational symmetry and smoothness of the
projected spatially periodic orbits. Both of these features are
unattainable in curved, two-dimensional space-times, and hence can
be reasonably waived.

\section{Revised wording of the cosmological analogy}

In view of the above discussion, the wording of the mathematical
analogy between coherent Lagrangian eddy boundaries and photon
spheres in cosmology \citep[][p.\ 731, para.\ 2]{Haller-Beron-13}
should be revised as follows:
\begin{quotation}
  The closed $\lambda$-lines we have been seeking are therefore
  closed null-geodesics of $g_{\lambda}$. Trajectories spanning
  these geodesics project along the local time-like coordinates
  (trajectories of the $\xi_{1}$ vector field) onto periodic
  trajectories on the local space-like coordinates (trajectories
  of the $\xi_{2}$ vector field). In cosmology, the projection of
  some families of null-geodesics from space-time onto the space-like
  variables also produces closed trajectories. Such families of
  null-geodesics are often referred to as \emph{photon spheres},
  as their spatial projections trap photons orbiting around black
  holes \citep{Claudel-etal-01}. In the cosmological context,
  null-geodesics are tangent to \emph{light cones}, which in our
  case are formed by the two vectors $\eta_{\lambda}^{\pm}(x_{0})$
  at $x_{0}$ (Figure 2 of Haller \& Beron-Vera 2013).
\end{quotation}

We acknowledge comments by Angus Prain and Valerio Faraoni that
made us realize the need for this clarifying note.  We are grateful
to Mohammad Farazmand for his numerical work in connection with
Figure 1. FJBV acknowledges support by NSF grant CMG0825547 and by
a grant from the BP/Gulf of Mexico Research Initiative.

\bibliographystyle{jfm}

\end{document}